\documentclass[11pt]{article}
\usepackage{amsfonts,epsfig,latexsym}

\newcommand{\R}{{\mathbb{R}}}
\newcommand{\Z}{{\mathbb{Z}}}

\newcommand{\C}{{\mathbb{C}}}
\newcommand{\I}{{\mathbb{I}}}
\newcommand{\beq}{\begin{equation}}
\newcommand{\eeq}{\end{equation}}
\newcommand{\bea}{\begin{eqnarray}}
\newcommand{\eea}{\end{eqnarray}}
\newcommand{\ra}{\rightarrow}

\newcommand{\thet}{\vartheta}
\newcommand{\fl}{{\it Fl}}
\newcommand{\err}{{\rm Err}}
\newcommand{\tr}{{\rm tr}}

\begin{document}

\title{Instability of breathers in the topological discrete sine-Gordon 
system}
\author{
J.M. Speight\thanks{E-mail: {\tt j.m.speight@leeds.ac.uk}}\\
Department of Pure Mathematics, University of Leeds\\
Leeds LS2 9JT, England}

\date{}

\maketitle
\begin{abstract}
It is demonstrated that the breather solutions recently discovered in the
weakly coupled topological discrete sine-Gordon system are spectrally
unstable. This is in contrast with more conventional 
spatially discrete systems, whose breathers are always spectrally stable
at sufficiently weak coupling.
\end{abstract}

\section{Introduction}
\label{int}

There has long been interest in solitonic field theories defined on 
spacetimes where time is continuous but space is a discrete lattice.
The motivation for this scenario varies from physical reality
(condensed matter and biophysics, see \cite{condmat} and references
therein), through
mathematical convenience
(integrable quantum field theory \cite{fadtak}) 
to computational necessity (when 
numerically simulating soliton dynamics on a computer). One particularly
interesting programme, whose motivation is of the last type, is that of
Ward. The aim here is to find spatially discrete versions of Bogomol'nyi
field theories which preserve the Bogomol'nyi properties of their continuum
counterparts: a topological lower bound on soliton energy saturated by
solutions of a first order ``self-duality'' equation, these solutions
occuring in continuous families, or ``moduli spaces.'' 
Such systems appear to capture the qualitative features of the continuum
soliton dynamics even on very coarse lattices, 
making them ideal candidates for computer
simulations.
The programme is
essentially complete for one spatial dimension, that is, such a 
``topological'' discretization has been constructed for any kink-bearing
nonlinear Klein-Gordon model \cite{topkink}. Results in higher dimensions
are more partial \cite{ward,theod,lees}.

The most heavily studied example is the topological discrete sine-Gordon
(TDSG) system \cite{tdsg} and its variants \cite{zak}. Here one has an
explicit, {\em continuous\, } one-dimensional moduli space of static kink
solutions, despite the discrete nature of space, and the kink dynamics is
similar to continuum kink dynamics even when the kink structure covers only
2 or 3 lattice sites. This should be contrasted with the conventional
discretization, the Frenkel-Kontorova model, in which the kink dynamics is
extremely complicated \cite{peykru}. 

Kinks and antikinks are not the only solitons of the continuum SG equation.
It also supports {\em breathers}, that is, spatially localized, time
periodic, oscillatory solutions. It was recently proved \cite{hasspe} that
the TDSG system also supports breathers, provided the lattice is sufficiently
coarse. So the question arises: to what extent are these breathers 
similar to their continuum counterparts? In this note we will demonstrate
one crucial difference: discrete breathers are spectrally unstable.

\section{Breathers in the TDSG system}
\label{bre}

The TDSG system is defined on spacetime $\Z\times\R$. Space is a regular
discrete lattice of spacing $h\in (0,2]$ and spatial position will be denoted
by $n\in\Z$. Time is continuous, denoted $t$. The field $\psi_n(t)$
evolves according to the differential-difference equation
\beq
\label{1}
\ddot{\psi}_n=\frac{4-h^2}{4h^2}\cos\psi_n(\sin\psi_{n+1}+\sin\psi_{n-1})
-\frac{4+h^2}{4h^2}\sin\psi_n(\cos\psi_{n+1}+\cos\psi_{n-1}).
\eeq
Note that this tends, as $h\ra 0$, to the usual sine-Gordon PDE (in 
laboratory coordinates $(x=nh,t)$) for $\phi=2\psi$. Equation (\ref{1})
supports a continuous translation orbit of static kink solutions,
\beq
\label{2}
\psi_n=2\tan^{-1}e^{a(n-b)},\qquad b\in\R
\eeq
where $a=\ln[(2+h)/(2-h)].$ The conserved energy associated with (\ref{1}),
call it $E$, satisfies $E\geq 1$ for all $\psi$ satisfying kink boundary
conditions ($\psi_n\ra 0$, $n\ra-\infty$; $\psi_n\ra\pi$, $n\ra\infty$),
and the static solutions (\ref{2}) saturate this bound. Note that as $h\ra 2$
the kinks tend to step functions, which is why one insists $h\leq 2$.

Breather solutions to (\ref{1}) of any period $T>2\pi$ where $T/2\pi\notin
\Z$, arise as follows. Consider the case $h=2$. Then (\ref{1}) reduces to
\beq
\label{3}
\ddot{\psi}_n=-\frac{1}{2}\sin\psi_n(\cos\psi_{n+1}+\cos\psi_{n-1}),
\eeq
which supports \cite{tdsg} the ``one-site breather''
\beq
\label{4}
\Psi_n(t)=\left\{\begin{array}{cl}
0 & n\neq 0 \\
\thet(t) & n=0,
\end{array}\right.
\eeq
where $\thet(t)$ is any $T$-periodic solution of the pendulum equation
\beq
\label{5}
\ddot{\thet}+\sin\thet=0.
\eeq
By applying an implicit function theorem argument, in the manner of
MacKay and Aubry \cite{macaub}, one obtains the following \cite{hasspe}:
For all $T>2\pi$, $T\notin 2\pi\Z$, there exists a continuous deformation
of the one-site breather (\ref{4}) through breather solutions of (\ref{1})
with lattice spacing $h\in[2-\epsilon,2]$, where $\epsilon>0$ is sufficiently
small. So breather solutions exist for all $h$ sufficiently close to 2,
provided the breather period is nonresonant (not a multiple of $2\pi$).

By truncating the system to $N=2m+1$ sites ($-N\leq n\leq N$) and applying
a Newton-Raphson scheme to $P_T-{\rm Id}$, where $P_T:\R^{2N}\ra\R^{2N}$
is the period $T$ Poincar\'e return map for system (\ref{1}), one can 
construct these breather solutions numerically \cite{aubmar}. One starts at
$h=2-\delta h$, $\delta h$ small, with the one-site breather as an initial
guess, then interates the NR scheme until $P_T-{\rm Id}$ converges (within
some tolerance) to 0. Repeating this for $h=2-2\delta h$ with the newly found
breather as initial guess, one works piecemeal away from the $h=2$ limit.
In this way a portrait of the existence domain of breathers in the
$(T,h)$ parameter space has been built up \cite{hasspe}. 

\section{Spectral instability}
\label{spec}

One says that a period $T$ solution $q(t)$ of a dynamical system on phase
space $M$ is spectrally stable if the spectrum of the associated Floquet
map $\fl:T_{q(0)}M\ra T_{q(T)}M=T_{q(0)}M$ is contained within the
closed unit disk
$D\subset\C$. Here $\fl=DP_T$, that is the period $T$ return map of the
{\em linearized\, } flow about $q$. Spectral stability is necessary, but
not sufficient, for linear and hence practical stability. We shall show that
the one-site breather (\ref{4}) is a spectrally {\em unstable\, } solution
of (\ref{3}). Since the spectrum of $\fl$ varies continuously under
constinuous perturbation, it follows that all breathers, for $h$ sufficiently
close to $2$, are unstable.

Let $\psi_n(t)$ be a solution of (\ref{1}), and $\delta\psi_n(t)$ be a 
solution of the linearized flow about $\psi_n(t)$. Then
\bea
\delta\ddot{\psi}_n&=&
-[\frac{4-h^2}{4h^2}\sin\psi_n(\sin\psi_{n+1}+\sin\psi_{n-1})
+\frac{4+h^2}{4h^2}\cos\psi_n(\cos\psi_{n+1}+\cos\psi_{n-1})]
\delta\psi_n \nonumber \\
& &+
\frac{4-h^2}{4h^2}\cos\psi_n(\cos\psi_{n+1}\delta\psi_{n+1}
+\cos\psi_{n-1}\delta\psi_{n-1}) \nonumber \\
\label{6}
& &
+
\frac{4+h^2}{4h^2}\sin\psi_n(\sin\psi_{n+1}\delta\psi_{n+1}
+\sin\psi_{n-1}\delta\psi_{n-1}).
\eea
In particular, if $h=2$ and 
$\psi_n(t)$ is the one-site breather (\ref{4}), one finds
that
\beq
\label{7}
\delta\ddot{\psi}_n=\left\{
\begin{array}{cl}
-\delta\psi_n & |n|>1 \\
-\frac{1}{2}(1+\cos\thet)\delta\psi_n & |n|=1 \\
-\cos\thet\, \delta\psi_n & n=0.
\end{array}\right.
\eeq
Truncating the lattice to size $N=2m+1$ (our result is independent of $m$),
the Floquet map takes the initial data of $\delta\psi$ to the final data
(at $t=T$),
\beq
\label{8}
\fl:\left[\begin{array}{c}
\delta\psi_{-m}(0) \\
\delta\dot{\psi}_{-m}(0) \\
\vdots \\
\delta\psi_{m}(0) \\
\delta\dot{\psi}_{m}(0)
\end{array}\right]\mapsto
\left[\begin{array}{c}
\delta\psi_{-m}(T) \\
\delta\dot{\psi}_{-m}(T) \\
\vdots \\
\delta\psi_{m}(T) \\
\delta\dot{\psi}_{m}(T)
\end{array}\right]
\eeq
Of course, $\fl$ is linear by linearity of (\ref{7}). In fact, since
(\ref{1}) defines a Hamiltonian flow, $\fl$ is
symplectic with respect to the natural symplectic structure on $\R^{2N}$, 
that is
\beq
\label{8.5} 
\fl^T\Omega\fl=\Omega,
\eeq
where $\Omega$ is the block diagonal matrix
\beq
\label{8.55}
\Omega={\rm diag}(J,\ldots,J),\qquad J=\left(\begin{array}{cc}
0 & 1 \\ -1 & 0 \end{array}\right).
\eeq
Note that $\fl$ is continuously connected through $GL(2N,\R)$ to $\I_{2N}$
(by the time evolution (\ref{6})), so (\ref{8.5}) implies $\det\fl=1$.
Symplectomorphicity imposes many more constraints on the spectrum of $\fl$
\cite{aub}, but we shall not need these.

Since system (\ref{7}) is decoupled, $\fl$ is block diagonal with one
symplectic
$2\times 2$ block $\fl_n$ per lattice site. To demonstrate instability, 
therefore, it suffices to exhibit one block with an eigenvalue outside
$D=\{z\in\C:|z|\leq 1\}$. The offending block is $\fl_1$ (or $\fl_{-1}$):
\beq
\label{9}
\fl_1=\left[\begin{array}{cc}
y_1(T) & y_2(T) \\
\dot{y}_1(T) & \dot{y}_2(T) 
\end{array}\right]
\eeq
where $y_1,y_2$ form a basis of solutions of
\beq
\label{10}
\ddot{y}+\frac{1}{2}(1+\cos\thet(t))y=0
\eeq
with $y_1(0)=\dot{y}_2(0)=1$, $\dot{y}_1(0)=y_2(0)=0$. Note that, as argued 
above, $\det\fl_1=1$, so if
$|{\rm tr}\, \fl_1|>2$, the eigenvalues are a real conjugate pair
$\{\lambda,1/\lambda\}$ with $|\lambda|>1$, and $\Psi_n(t)$ is spectrally
unstable.

It is numerically trivial to solve (\ref{10}) (coupled to the pendulum
equation (\ref{5}) for $\thet(t)$) over $[0,T]$ and hence compute
${\rm tr}\, \fl_1$. The results are shown in figure 1.\, Clearly
${\rm tr}\, \fl_1>2$ for all $T$ within our range (the limit $T\ra\infty$
is numerically inaccessible), so both $\fl_1$ and $\fl_{-1}$ have a real eigenvalue exceeding 1, confirming our claim that the
breathers are unstable for $h$ close to $2$.

It is possible, of course, that as $h$ decreases substantially below 2, the
two offending  eigenvalues move
into $D$, so that spectral stability is recovered. To eliminate this 
possibility, we have computed the spectrum of $\fl$ at a sample of points
within the breather existence domain. This domain, whose construction is 
described in detail in \cite{hasspe}, is depicted in $(\omega,h)$ 
space ($\omega=2\pi/T$ being breather frequency) in figure 2, with the 
sampled points marked by crosses. Direct computation of the spectrum of $\fl$
is certainly a much cruder approach to numerical spectral stability analysis
than that of Aubry \cite{aub}. However, since we seek only to confirm
{\em instability}, rather than examine bifurcations between stability and
instability, a simple approach is justified.

At each point sampled, $\fl$ was constructed by solving (\ref{6}) 
approximately using a 4th order Runge-Kutta scheme on an 11 site lattice.
Note that no decoupling occurs now that $h\neq 2$.
Numerical accuracy of $\fl$ was tested by measuring the deviation of $\fl$
from symplectomorphicity by comparing the euclidean norm $||\err||$ of
\beq
\err=\fl^T\Omega\fl-\Omega,\qquad
||M||:=(\sum_{i,j}M_{ij}^2)^\frac{1}{2},
\eeq
with $||\fl||$. In each case it was found that the largest eigenvalue of
$\fl$ substantially exceeds 1. The results are summarized in table 
\ref{tab1}.

\section{Conclusion}

This paper reinforces the point that breathers in the weakly coupled TDSG
system are really quite different from continuum sine-Gordon breathers. Their
domain of existence does not extend to the continuum limit, their initial
profiles do not resemble continuum breathers sampled on a lattice 
\cite{hasspe}, and they are spectrally unstable. In this last property, they
are also different from breathers constructed in conventional discrete
systems (oscillator chains), which are ubiquitously stable, at least on
sufficiently coarse lattices \cite{aub}. 
We conclude that, while the TDSG system captures the dynamics of kinks and
antikinks quite faithfully even for large $h$, the same is  not true of
kink-antikink bound states, that is, breathers.  

\begin{table}
\label{tab1}
\begin{center}
\begin{tabular}{|c|c|c|c|c||c|c|c|c|c|} \hline
$\omega$ & $h$ & $||\err||$ & $||\fl||$ & $|\lambda|_{\max}$ &
$\omega$ & $h$ & $||\err||$ & $||\fl||$ & $|\lambda|_{\max}$ 
\\ \hline
0.40 & 1.83 & 0.0743 & 28.6358 & 45.2109 & 0.65 & 1.89 & 0.0062 & 4.4025 & 4.8864 \\
0.40 & 1.90 & 0.0854 & 22.3127 & 30.8483 & 0.70 & 1.79 & 0.0053 & 3.9849 & 3.9052 \\
0.45 & 1.87 & 0.0916 & 15.7065 & 21.7614 & 0.70 & 1.92 & 0.0004 & 3.9330 & 3.5739 \\
0.45 & 1.95 & 0.0084 & 10.6190 & 14.7902 & 0.75 & 1.83 & 0.0031 & 3.1875 & 2.9886 \\
0.55 & 1.85 & 0.0011 & 7.6868 & 10.1095 & 0.75 & 1.90 & 0.0013 & 3.0531 & 2.8147 \\
0.60 & 1.75 & 0.0040 & 5.5526 & 6.7439 & 0.80 & 1.88 & 0.0008 & 2.1615 & 2.2982 \\
0.60 & 1.82 & 0.0144 & 5.5808 & 7.3534 & 0.80 & 1.96 & 0.0006 & 2.1080 & 2.1469 \\
0.65 & 1.78 & 0.0040 & 4.4930 & 4.8864 & 0.90 & 1.95 & 0.0002 & 2.6793 & 1.4589 \\ \hline

\end{tabular}
\end{center}
\caption{ The largest eigenvalue of $\fl$ for various 
breathers.}
\end{table}

\section*{Acknowledgment}
The author is an EPSRC Postdoctoral Research Fellow in Mathematics.

\newpage
\section*{Figure captions}

Figure 1: Plot of ${\tr \fl_1}$ against breather period $T$. Note that
${\tr \fl_1}>2$ for all $T$.

\vspace{0.5cm}
\noindent
Figure 2: Domain of existence of discrete breathers in $(\omega,h)$ space.
The jagged dotted curve is the theoretical lower boundary obtained by a
phonon resonance argument. The solid curve is the actual lower boundary,
generated numerically as in \cite{hasspe}. The small $\omega$
region (effectively
$\omega<1/3$) is numerically inaccessible. Crosses mark those breathers
whose spectral instability was confirmed numerically (see table 1).

\newpage

\section*{Figure 1}

\vbox{
\centerline{\epsfysize=4truein
\epsfbox[55 178 570 615]{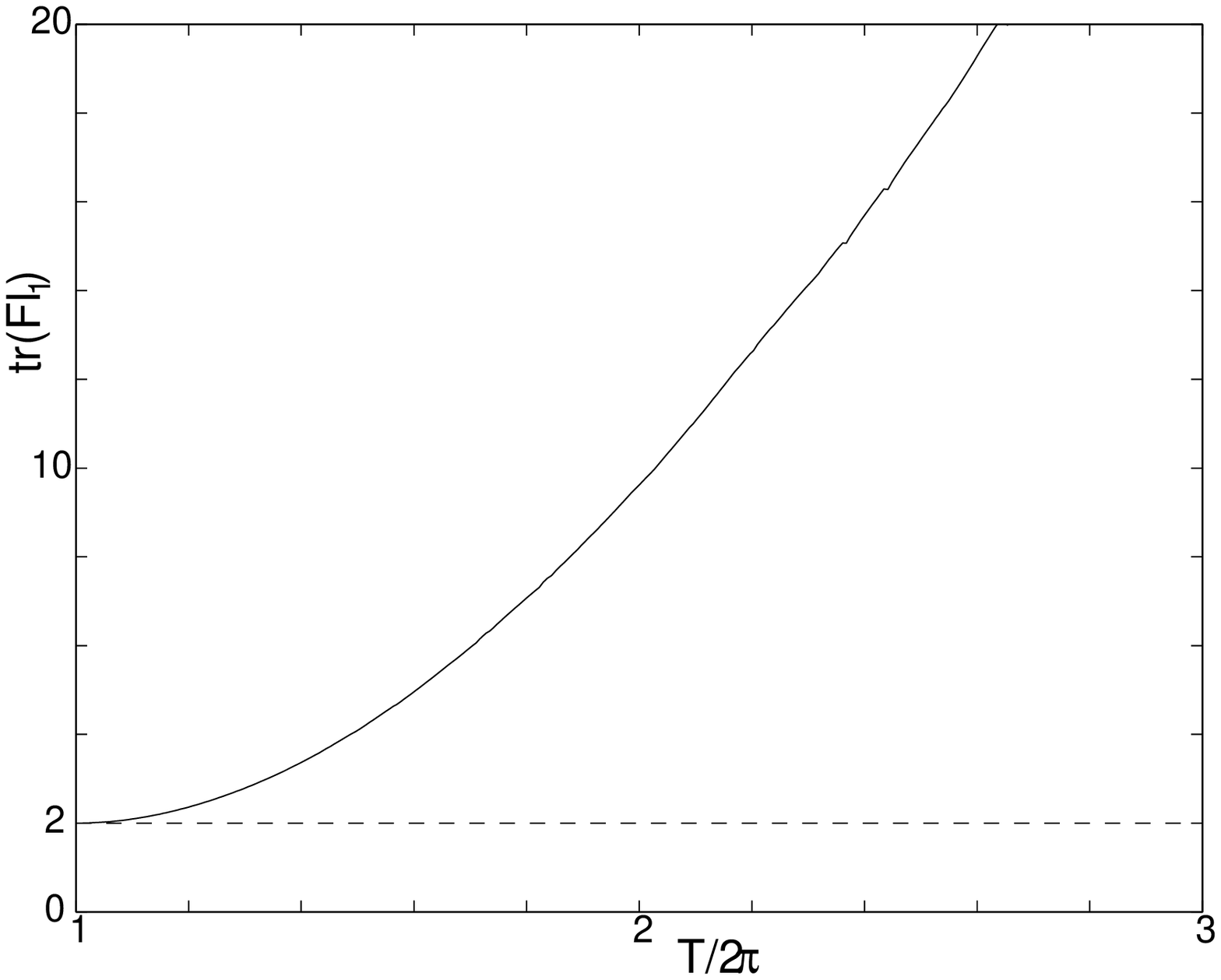}}
\noindent
}

\newpage

\section*{Figure 2}

\vbox{
\centerline{\epsfysize=4truein
\epsfbox[61 185 566 612]{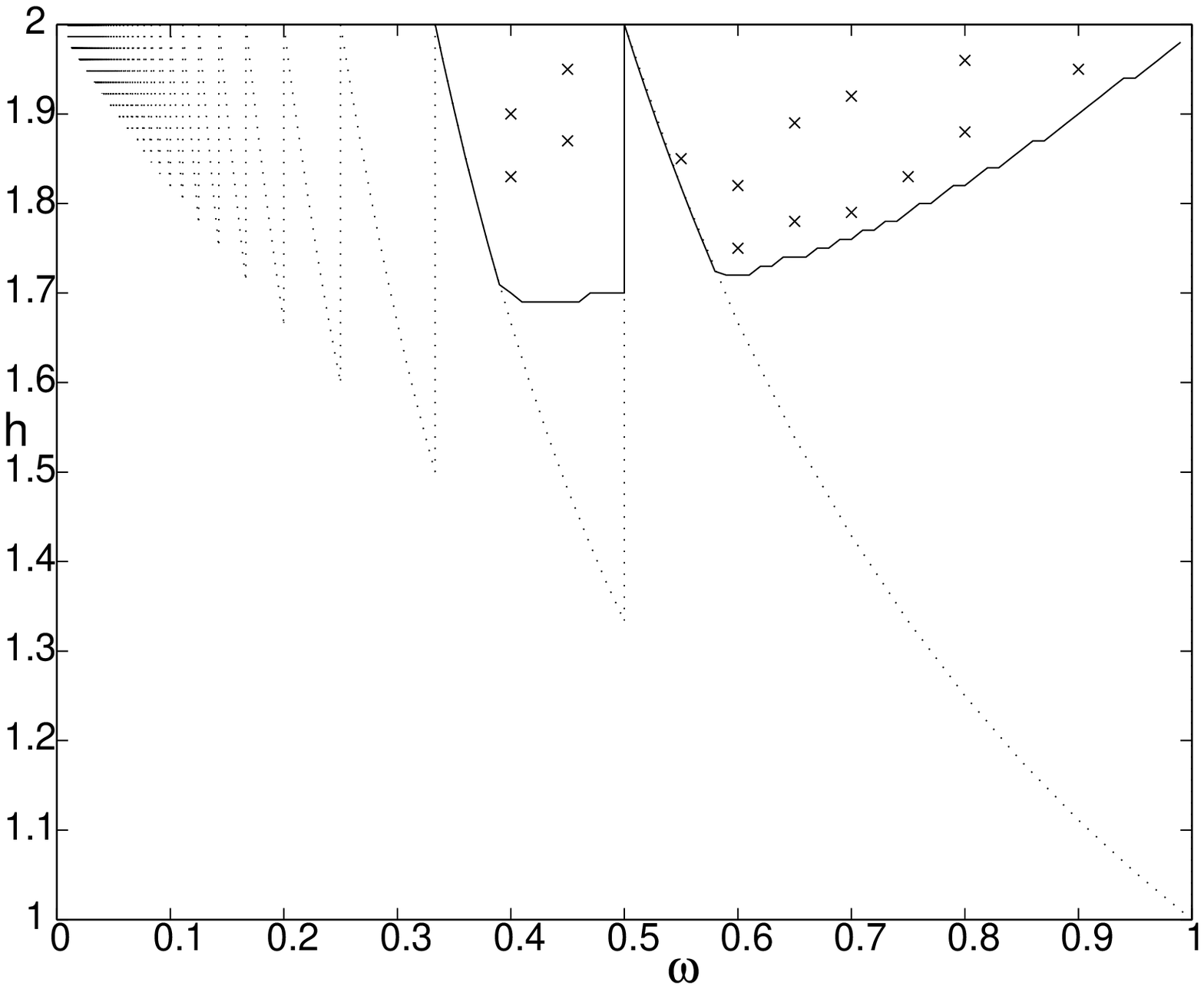}}
\noindent
}

\vspace{0.5cm}


\begin{thebibliography}{99}

\bibitem{condmat} R. Boesch, C.R. Willis and M. El-Batanouny,
``Spontaneous emission of radiation from a discrete sine-Gordon kink''
{\sl Phys.\ Rev.} {\bf B40} (1989) 2284.

\bibitem{fadtak} L.D. Faddeev and L.A. Takhtajan,
{\sl Hamiltonian Methods in the Theory of Solitons}
(Springer, London, 1987).

\bibitem{topkink} J.M. Speight,
``Topological discrete kinks''
{\sl Nonlinearity} {\bf 12} (1999) 1373.

\bibitem{ward} R.S. Ward,
``Bogomol'nyi bounds for two-dimensional lattice systems''
{\sl Commun.\ Math.\ Phys.} {\bf 184} (1997) 397.

\bibitem{theod} T. Ioannidou,
``Soliton dynamics in a novel discrete $O(3)$ sigma model in $(2+1)$
dimensions''
{\sl Nonlinearity} {\bf 10} (1997) 1357.

\bibitem{lees} R. Leese,
``Discrete Bogomol'nyi equations for the nonlinear $O(3)$ sigma model 
in $(2+1)$ dimensions''
{\sl Phys.\ Rev.} {\sl D40} (1989) 2004.

\bibitem{tdsg} J.M. Speight and R.S. Ward,
``Kink dynamics in a novel discrete sine-Gordon system''
{\sl Nonlinearity} {\bf 7} (1994) 475.

\bibitem{zak} W.J. Zakrzewski,
``A modified discrete sine-Gordon system''
{\sl Nonlinearity} {\bf 8} (1995) 517.

\bibitem{peykru} M. Peyrard and M.D. Kruskal,
``Kink dynamics in the highly discrete sine-Gordon system''
{\sl Physica} {\bf D14} (1984) 88.

\bibitem{hasspe} M. Haskins and J.M. Speight,
``Breathers in the weakly coupled topological discrete sine-Gordon system''
{\sl Nonlinearity} {\bf 11} (1998) 1651.

\bibitem{macaub} R.S. MacKay and S. Aubry,
``Proof of existence of breathers for time-Reversible or Hamiltonian networks
of weakly coupled oscillators''
{\sl Nonlinearity} {\bf 7} (1994) 1623.

\bibitem{aubmar} J.L. Mar\'{\i}n and S. Aubry,
``Breathers in nonlinear lattices: numerical calculation from the 
anticontinuous limit''
{\sl Nonlinearity} {\bf 9} (1996) 1501.

\bibitem{aub} S. Aubry,
``Breathers in nonlinear lattices: existnce, stability and quantization''
{\sl Physica} {\bf D103} (1997) 201.


\end{thebibliography}
\end{document}